\documentclass[twocolumn]{IEEEtran}
\usepackage{amsbsy}
\usepackage{latexsym}
\usepackage{amsmath}
\usepackage{times}
\input epsf
\title{Blind Adaptive Algorithms for Decision Feedback DS-CDMA Receivers
in Multipath Channels}

\author{Rodrigo C. de Lamare, IEEE Member and Raimundo Sampaio-Neto  \\
\thanks{This work was supported by the Brazilian Council for Scientific and
Technological Development (CNPq). Dr. R. C. de Lamare was with
CETUC/PUC-RIO and is now a Lecturer with the Communications
Research Group, Department of Electronics, University of York,
York Y010 5DD, United Kingdom and Prof. R. Sampaio-Neto is with
CETUC/PUC-RIO, 22453-900, Rio de Janeiro, Brazil. Phone:
+55-21-31141701 Fax: +55-21-22945748. E-mails:
rcdl500@ohm.york.ac.uk, raimundo@cetuc.puc-rio.br} }

\begin{document}
\maketitle
\thispagestyle{empty}
\begin{abstract}
{ In this work we examine blind adaptive and iterative decision
feedback (DF) receivers for direct sequence code division multiple
access (DS-CDMA) systems in frequency selective channels.
Code-constrained minimum variance (CMV) and constant modulus (CCM)
design criteria for DF receivers based on constrained optimization
techniques are investigated for scenarios subject to multipath.
Computationally efficient blind adaptive stochastic gradient (SG)
and recursive least squares (RLS) algorithms are developed for
estimating the parameters of DF detectors along with successive,
parallel and iterative DF structures. A novel successive parallel
arbitrated DF scheme is presented and combined with iterative
techniques for use with cascaded DF stages in order to mitigate
the deleterious effects of error propagation. Simulation results
for an uplink scenario assess the algorithms, the blind adaptive
DF detectors against linear receivers and evaluate the effects of
error propagation of the new cancellations techniques against
previously reported approaches.}

\end{abstract}

\section{Introduction}

\PARstart{C}{ode} division multiple access (CDMA) implemented with
direct sequence (DS) spread-spectrum signalling is amongst the
most promising multiple access technologies for current and future
communication systems. Such services include third-generation
cellular telephony, indoor wireless networks, terrestrial and
satellite communication systems. The advantages of CDMA include
good performance in multi-path channels, flexibility in the
allocation of channels, increased capacity in bursty and fading
environments and the ability to share bandwidth with narrowband
communication systems without deterioration of either's systems
performance \cite{1,verdu}.

Demodulating a desired user in a DS-CDMA network requires
processing the received signal in order to mitigate different
types of interference, namely, narrowband interference (NBI),
multi-access interference (MAI), inter-symbol interference (ISI)
and the noise at the receiver. The major source of interference in
most CDMA systems is MAI, which arises due to the fact that users
communicate through the same physical channel with non-orthogonal
signals. The conventional (single-user) receiver that employs a
filter matched to the signature sequence does not suppress MAI and
is very sensitive to differences in power between the received
signals (near-far problem). Multiuser detection has been proposed
as a means to suppress MAI, increasing the capacity and the
performance of CDMA systems \cite{1,verdu}. The optimal multiuser
detector of Verdu \cite{verdu86} suffers from exponential
complexity and requires the knowledge of timing, amplitude and
signature sequences. This fact has motivated the development of
various sub-optimal strategies: the linear \cite{lupas} and
decision feedback \cite{falconer} receivers, the successive
interference canceller \cite{patel} and the multistage detector
\cite{varanasi}. For uplink scenarios, decision feedback
detection, which is relatively simple and performs linear
interference suppression followed by interference cancellation was
shown to provide substantial gains over linear detection
\cite{falconer,duel,woodward1,woodward2}.

When used with short code or repeated spreading codes, adaptive
signal processing methods are suitable to CDMA systems because
they can track the highly dynamic conditions often encountered in
such systems due to the mobility of mobile terminals and the
random nature of the channel access. Adaptive techniques can also
alleviate the computational complexity required for parameter
estimation. In particular, blind adaptive signal processing is an
interesting alternative for situations where a receiver loses
track of the desired user and/or a training sequence is not
available. In this context, blind linear receivers for DS-CDMA
have been proposed in the last years to supress MAI
\cite{honig,miguez,xu&tsatsanis,kwak,tugnait,xu&liu}. Blind linear
solutions for flat channels have been reported for the first time
in \cite{honig}, where the blind detector was designed on the
basis of the minimum output energy (MOE) or minimum variance (MV).
Following the initial success of the MV receiver \cite{honig},
blind receivers using the constant modulus (CM) criterion, which
outperformed their MV counterparts, were reported in
\cite{miguez,kwak} and \cite{tugnait}. In this context, the work
by Tugnait and Li \cite{tugnait} is an inverse filtering criterion
and does not exploit the energy contained in the signal copies
available in multipath, leading to performance degradation as
compared to supervised solutions. In order to improve performance
and close the gap between blind and trained solutions, Xu and
Tsatsanis \cite{xu&tsatsanis} exploited the multipath components
through a constrained MV (CMV) method \cite{xu&tsatsanis} that
treats different signal copies as variables and jointly optimizes
the receiver and channel parameters. Another solution that
outperforms the CMV technique of \cite{xu&tsatsanis} was proposed
by Xu and Liu \cite{xu&liu} for multipath environments, in which
constrained adaptive linear receivers are derived based upon the
joint optimization of channel and receiver parameters in
accordance with the constant modulus criterion. Recently, a
code-constrained CM design for linear receivers and an RLS
algorithm, that outperform previous approaches, were presented in
\cite{delamarevtc} for a downlink scenario.

Although relatively simple, DF structures can perform
significantly better than linear systems and the existing work on
blind adaptive DF receivers was restricted to single-path channels
solutions \cite{ping,choi,dianat} and have to be modified for
multipath. Detectors with DF are especially interesting because
they offer the possibility of different types of cancellation,
namely, successive \cite{duel,varanasi2}, parallel
\cite{woodward1} and iterative \cite{woodward2,woodward3}, which
lead to different performances and degrees of robustness against
error propagation. This paper addresses blind adaptive DF
detection for multipath channels in DS-CDMA systems based on
constrained optimization techniques using the MV and CM criteria.
The CMV and CCM solutions for the design of blind DF CDMA
receivers are presented and then computationally efficient blind
adaptive algorithms are developed for MAI, intersymbol
interference (ISI) suppression and channel estimation. The second
contribution of this work is a novel successive parallel
arbitrated DF structure based on the recent concept of parallel
arbitration \cite{barriac}. The new DF detector is then combined
with iterative cascaded DF stages, resulting in an improved DF
receiver structure that is compared with previously reported
methods. Computer simulations experiments show the effectiveness
of the proposed blind DF system for refining soft estimates and
mitigating the effects of error propagation.

This paper is organized as follows. Section II briefly describes
the DS-CDMA communication system model. The constrained decision
feedback receivers and the blind channel estimation procedure are
described in Section III. Section IV is devoted to the successive
parallel arbitrated and iterative DF cancellation techniques,
whereas Section V is dedicated to the derivation of adaptive SG
algorithms and RLS type algorithms. Section VI presents and
discusses the simulation results and Section VII gives the
conclusions of this work.

\section{DS-CDMA system model}

Let us consider the uplink of a symbol synchronous binary
phase-shift keying (BPSK) DS-CDMA system with $K$ users, $N$ chips
per symbol and $L_{p}$ propagation paths. It should be remarked
that a synchronous model is assumed for simplicity, although it
captures most of the features of more realistic asynchronous
models with small to moderate delay spreads. The baseband signal
transmitted by the $k$-th active user to the base station is given
by {
\begin{equation}
x_{k}(t)=A_{k}\sum_{i=-\infty}^{\infty}b_{k}(i)s_{k}(t-iT)\end{equation}
} where $b_{k}(i) \in \{\pm1\}$ denotes the $i$-th symbol for user
$k$, the real valued spreading waveform and the amplitude
associated with user $k$ are $s_{k}(t)$ and $A_{k}$, respectively.
The spreading waveforms are expressed by { $s_{k}(t) =
\sum_{i=1}^{N}a_{k}(i)\phi(t-iT_{c})$}, where { $a_{k}(i)\in
\{\pm1/\sqrt{N} \}$}, { $\phi(t)$} is the chip waverform, $T_{c}$
is the chip duration and $N=T/T_{c}$ is the processing gain.
Assuming that the receiver is synchronised with the main path, the
coherently demodulated composite received signal is
\begin{equation}
r(t)= \sum_{k=1}^{K}\sum_{l=0}^{L_{p}-1}
h_{k,l}(t)x_{k}(t-\tau_{k,l})+n(t)
\end{equation}
where $h_{k,l}(t)$ and $\tau_{k,l}$ are, respectively, the channel
coefficient and the delay associated with the $l$-th path and the
$k$-th user. Assuming that $\tau_{k,l} = lT_{c}$, the channel is
constant during each symbol interval and the spreading codes are
repeated from symbol to symbol, the received signal $ r(t)$ after
filtering by a chip-pulse matched filter and sampled at chip rate
yields the $M$-dimensional received vector
\begin{equation}
{\bf r}(i) = \sum_{k=1}^{K} {\bf H}_{k}(i) A_{k} {\bf S}_{k}{\bf
b}_{k}(i) + {\bf n}(i)
\end{equation}
where $M=N+L_{p}-1$, ${\bf n}(i) = [n_{1}(i)
~\ldots~n_{M}(i)]^{T}$ is the complex Gaussian noise vector with
$E[{\bf n}(i){\bf n}^{H}(i)] = \sigma^{2}{\bf I}$, where
$(\cdot)^{T}$ and $(\cdot)^{H}$ denote transpose and Hermitian
transpose, respectively, $E[\cdot]$ stands for ensemble average,
the user symbol vector is ${\bf b}_{k}(i) = [b_{k}(i)~\ldots ~
b_{k}(i-L_{s}+1)]^{T}$, the amplitude of user $k$ is $A_{k}$, the
channel vector of user $k$ is { ${\bf h}_{k}(i) = [h_{k,0}(i)
\ldots h_{k,L_{p}-1}(i)]^{T}$}, $L_{s}$ is the ISI span and the {
$(L_{s}\times N)\times L_{s}$} diagonal matrix { ${\bf S}_{k}$}
with $N$-chips shifted versions of the signature of user $k$ is
given by
\begin{equation}
{\bf S}_{k} = \left[\begin{array}{c c c c c c c}
{\bf s}_{k} & {\bf 0} & \ldots & {\bf 0}  \\
 {\bf 0} & {\bf s}_{k} & \ddots  &  {\bf 0} \\
 \vdots & \vdots & \ddots & \vdots \\
 {\bf 0} & {\bf 0} & \ldots & {\bf s}_{k} \end{array}\right]
 \end{equation}
where ${\bf s}_{k} = [a_{k}(1) \ldots a_{k}(N)]^{T}$ is the
signature sequence for the $k$-th user and the { $ M~\times
(L_{s}\times N)$} channel matrix { ${\bf H}_{k}(i)$} for user $k$
is
\begin{equation}
{\bf H}_{k}(i) = \left[\begin{array}{c c c c c c c }
h_{k,0}(i) &  \ldots & h_{k,L_{p}-1}(i) & \ldots & 0 & 0  \\
\vdots  & \ddots & \ddots & \ddots & \ddots & \vdots \\
 0 & 0 &  \ldots & h_{k,0}(i) & \ldots & h_{k,L_{p}-1}(i)
\end{array}\right]
\end{equation}
where $h_{k,l}(i) = h_{k,l}(iT_{c})$. The MAI comes from the
non-orthogonality between the received signature sequences,
whereas the ISI span $L_{s}$ depends on the length of the channel
response, which is related to the length of the chip sequence. For
{ $L_{p}=1,~ L_{s}=1$} (no ISI), for { $1<L_{p}\leq N, L_{s}=2$},
for { $N <L_{p}\leq 2N, L_{s}=3$}.

\section{Blind Decision Feedback Constrained Receivers}

\begin{figure}[!htb]
\begin{center}
\def\epsfsize#1#2{0.75\columnwidth}
\epsfbox{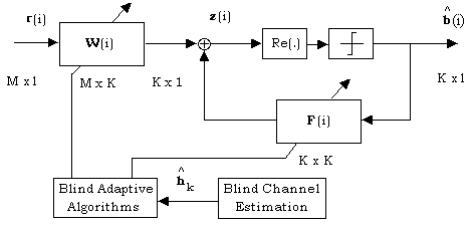} \caption{Block diagram of a blind multiuser
decision feedback receiver. }
\end{center}
\end{figure}

Let us describe the design of synchronous blind decision feedback
constrained detectors, as the one shown in Fig. 1. It should be
remarked that portions of the material presented here were
presented in \cite{delamareprc}. Consider the received vector
${\bf r}(i)$, and let us introduce the $M\times L_{p}$ constraint
matrix ${\bf C}_{k}$ that contains one-chip shifted versions of
the signature sequence for user $k$:
\begin{equation}
{\bf C}_{k} = \left[\begin{array}{c c c }
a_{k}(1) &  & {\bf 0} \\
\vdots & \ddots & a_{k}(1)  \\
a_{k}(N) &  & \vdots \\
{\bf 0} & \ddots & a_{k}(N)  \\
 \end{array}\right]\end{equation}
The input to the hard decision device, depicted in Fig. 1,
corresponding to the $i$th symbol is
\begin{equation}
{\bf z}(i) = {\bf W}^{H}(i){\bf r}(i) - {\bf F}^{H}(i) \hat{\bf
b}(i)
\end{equation}
where the input ${\bf z}(i) = [z_{1}(i) ~\ldots ~z_{K}(i)]^{T}$,
${\bf W}(i) = [{\bf w}_{1}~ \ldots~{\bf w}_{K}]$ is $M \times K$
the feedforward matrix, $\hat{\bf b}(i)=[b_{1}(i)
~\ldots~b_{K}(i)]^{T}$ is the $K \times 1$ vector of estimated
symbols, which are fed back through the $K \times K$ feedback
matrix ${\bf F}(i)=[ {\bf f}_{1}(i)~\ldots~{\bf f}_{K}(i)]$.
Generally, the DF receiver design is equivalent to determining for
user $k$ a feedforward filter ${\bf w}_{k}(i)$ with $M$ elements
and a feedback one ${\bf f}_{k}(i)$ with $K$ elements that provide
an estimate of the desired symbol:
\begin{equation}
z_{k}(i) = {\bf w}_{k}^{H}(i){\bf r}(i)-{\bf f}_{k}^{H}(i)\hat{\bf
b}(i)~, ~~~~~~k=1,2,\ldots,K
\end{equation}
where $\hat{\bf b}(i)=sgn[\Re({\bf W}^{H}(i){\bf r}(i))]$ is the
vector with initial decisions provided by the linear section, {
${\bf w}_{k}$} and ${\bf f}_{k}$ are optimized by the MV or the CM
cost functions, subject to a set multipath constraints given by {
${\bf C}_{k}^{H}{\bf w}_{k}(i) = {\bf h}_{k}(i)$} for the MV case,
or ${\bf C}_{k}^{H}{\bf w}_{k}(i) = \nu {\bf h}_{k}(i)$ for the CM
case, where $\nu$ is a constant to ensure the convexity of the
CM-based receiver and { ${\bf h}_{k}(i)$} is the $k$th user
channel vector. In particular, the feedback filter ${\bf
f}_{k}(i)$ of user $k$ has a number of non-zero coefficients
corresponding to the available number of feedback connections for
each type of cancellation structure. The final detected symbol is
obtained with:
\begin{equation}
 \hat{b}_{k}^{f}(i) = sgn\Big(\Re\Big[z_{k}(i)\Big]\Big) = sgn\Big(\Re\Big[{\bf w}_{k}^{H}(i){\bf
r}(i)-{\bf f}_{k}^{H}(i)\hat{\bf b}(i)\Big]\Big)~,
~~~~~~k=1,2,\ldots,K
\end{equation}
where $\Re(.)$ selects the real part and $sgn(.)$ is the signum
function. For successive DF (S-DF) \cite{duel}, the $K\times K$
matrix ${\bf F}(i)$ is strictly lower triangular, whereas for
parallel DF (P-DF) \cite{woodward1,woodward2} ${\bf F}(i)$ is full
and constrained to have zeros on the main diagonal in order to
avoid cancelling the desired symbols. The S-DF structure is
optimal in the sense of that it achieves the sum capacity of the
synchronous CDMA channel with AWGN \cite{varanasi2}. In addition,
the S-DF scheme is less affected by error propagation although it
generally does not provide uniform performance over the user
population, which is a desirable characteristic for uplink
scenarios. In this context, the P-DF system can offer uniform
performance over the users but is it suffers from error
propagation. In order to design the DF receivers and satisfy the
constraints of S-DF and P-DF structures, the designer must obtain
the vector with initial decisions $\hat{\bf b}(i)=sgn[\Re({\bf
W}^{H}(i){\bf r}(i))]$ and then resort to the following
cancellation approach. The non-zero part of the filter ${\bf
f}_{k}$ corresponds to the number of used feedback connections and
to the users to be cancelled. For the S-DF, the number of feedback
elements and their associated number of non-zero filter
coefficients in ${\bf f}_{k}$ (where $k$ goes from the second
detected user to the last one) range from $1$ to $K-1$. For the
P-DF, the feedback connections used and their associated number of
non-zero filter coefficients in ${\bf f}_{k}$ are equal to $K-1$
for all users and the matrix ${\bf F}(i)$ has zeros on the main
diagonal to avoid cancelling the desired symbols.

In what follows, constrained CM and MV design criteria for DF
detectors are presented. The CMV design for DF receivers
generalizes the work on linear structures of Xu and Tsatsanis
\cite{xu&tsatsanis}, whereas the CCM design is proposed here for
both linear and DF schemes.

\subsection{DF Constrained Constant Modulus (DF-CCM) Receivers}

To describe the DF-CCM receiver design let us consider the CM cost
function:
\begin{equation}
J_{CM}(i) = E\Big[(|{\bf w}_{k}^{H}(i){\bf r}(i) -{\bf
f}_{k}^{H}(i)\hat{\bf b}(i)|^{2}-1)^{2}\Big] = E[(|z_{k}(i)|^{2} -
1)^{2}]
\end{equation}
subject to ${\bf C}_{k}^{H}{\bf w}_{k}(i) = \nu {\bf h}_{k}(i)$,
where $z_{k}(i)={\bf w}_{k}^{H}(i){\bf r}(i)-{\bf
f}_{k}^{H}(i)\hat{\bf b}(i)$. Assuming that the channel vector
${\bf h}_{k}$ is known, let us consider the unconstrained cost
function ${J}_{CM}'(i) = E\Big[(|{\bf w}_{k}^{H}(i){\bf r}(i)
-{\bf f}_{k}^{H}(i)\hat{\bf b}(i)|^{2}-1)^{2}\Big] + 2 \Re[ ({\bf
C}_{k}^{H}{\bf w}_{k}(i)- \nu~{\bf
h}_{k}(i))^{H}\boldsymbol{\lambda}]$, where $\boldsymbol{
\lambda}$ is a vector of complex Lagrange multipliers. The
function ${J}_{CM}'(i)$ is minimized with respect to { ${\bf
w}_{k}(i)$} and { ${\bf f}_{k}(i)$} under the set of constraints {
${\bf C}_{k}^{H}{\bf w}_{k}(i) = \nu~{\bf h}_{k}(i)$}. Taking the
gradient terms of $J_{CM}(i)'$ with respect to ${\bf w}_{k}(i)$
and setting them to zero we have $\nabla J_{CM}(i)' = 2E[(|{\bf
w}_{k}^{H}(i){\bf r}(i)-{\bf f}_{k}^{H}(i)\hat{\bf b}(i)|^{2}-1)
{\bf r}(i)({\bf r}^{H}(i){\bf w}_{k}(i)-\hat{\bf b}^{H}(i){\bf
f}_{k}(i))] + 2{\bf C}_{k} \boldsymbol{\lambda} = {\bf 0}$, then
rearranging the terms we obtain $E[|z_{k}(i)|^{2}{\bf r}(i){\bf
r}^{H}(i)] {\bf w}_{k}(i) = E[z_{k}^{*}(i){\bf r}(i)] +
E[|z_{k}(i)|^{2}{\bf r}(i)\hat{\bf b}^{H}(i)]{\bf f}_{k}(i) - {\bf
C}_{k} \boldsymbol{\lambda}$ and consequently ${\bf w}_{k}(i) =
{\bf R}_{k}^{-1}(i)[{\bf d}_{k}(i) + {\bf T}_{k}(i){\bf f}_{k}(i)
- {\bf C}_{k} \boldsymbol{\lambda}]$, where ${\bf
R}_{k}(i)=E[|z_{k}(i)|^{2}{\bf r}(i){\bf r}^{H}(i)]$, ${\bf
T}_{k}(i) = E[|z_{k}(i)|^{2}{\bf r}(i)\hat{\bf b}(i)]$, ${\bf
d}_{k}(i)=E[z_{k}^{*}(i){\bf r}(i)]$ and the asterisk denotes
complex conjugation. Using the constraint ${\bf C}_{k}^{H}{\bf
w}_{k}(i) = \nu~{\bf h}_{k}(i)$ we arrive at the expression for
the Lagrange multiplier $\boldsymbol{\lambda} =({\bf
C}_{k}^{H}{\bf R}_{k}^{-1}(i){\bf C}_{k})^{-1} ({\bf
C}_{k}^{H}{\bf R}_{k}^{-1}(i){\bf T}_{k}(i){\bf f}_{k}(i) + {\bf
C}_{k}^{H}{\bf R}_{k}^{-1}(i){\bf d}_{k}(i) - \nu~{\bf
h}_{k}(i))$. By substituting $\boldsymbol{\lambda}$ into ${\bf
w}_{k}(i) = {\bf R}_{k}^{-1}(i)[{\bf d}_{k}(i) + {\bf
T}_{k}(i){\bf f}_{k}(i) - {\bf C}_{k} \boldsymbol{\lambda}]$ we
obtain the solution for the feedforward section of the DF-CCM
receiver:
\begin{equation}
{\bf w}_{k}(i) = {\bf R}_{k}^{-1}(i)\Bigg[{\bf d}_{k}(i) + {\bf
T}_{k}(i){\bf f}_{k}(i) - {\bf C}_{k} ({\bf C}_{k}^{H}{\bf
R}_{k}^{-1}(i){\bf C}_{k})^{-1} \times $$ $$ \Big({\bf
C}_{k}^{H}{\bf R}_{k}^{-1}(i){\bf T}_{k}(i){\bf f}_{k}(i) + {\bf
C}_{k}^{H}{\bf R}_{k}^{-1}(i){\bf d}_{k}(i) - \nu~{\bf
h}_{k}(i))\Big)\Bigg]
\end{equation}
where the expression in (11) is a function of previous values of
${\bf w}_{k}(i)$ and the channel ${\bf h}_{k}(i)$. To obtain the
CCM solution for the parameter vector ${\bf f}_{k}$ of the
feedback section, we compute the gradient terms of $J_{CM}'$ with
respect to ${\bf f}_{k}$ and by setting them to zero we have
$\nabla J_{CM}'(i) = 2 E[(|z_{k}(i)|^{2} -1) \hat{\bf b}(i) ({\bf
r}^{H}(i){\bf w}_{k}(i)-\hat{\bf b}^{H}(i){\bf f}_{k}(i))] = {\bf
0}$, then rearranging the terms we get $ E[|z_{k}(i)|^{2}\hat{\bf
b}(i)\hat{\bf b}^{H}(i)]{\bf f}_{k}(i) =E[|z_{k}(i)|^{2}\hat{\bf
b}(i){\bf r}^{H}(i)] {\bf w}_{k}(i) - E[z_{k}^{*}(i)\hat{\bf
b}(i)]$ and consequently we have
\begin{equation}
{\bf f}_{k}(i) = {\bf I}_{k}^{-1}\Big[{\bf T}_{k}^{H}(i){\bf
w}_{k}(i) - {\bf v}_{k}(i)\Big]
\end{equation}
where ${\bf I}_{k}=E[|z_{k}(i)|^{2}\hat{\bf b}(i)\hat{\bf
b}^{H}(i)]$ and ${\bf v}_{k} = E[z_{k}^{*}(i)\hat{\bf b}(i)]$. We
remark that (11) and (12) should be iterated in order to estimate
the desired user symbols.  The CCM linear receiver solution
proposed in \cite{delamarevtc} is obtained by making ${\bf
f}_{k}(i)= {\bf 0}$ in (11). An analysis of the CCM method in the
Appendix I examines its convergence properties for the linear
receiver case, extending previous results on its convexity for
both complex and multipath signals. Since the optimization of the
CCM cost function for a linear receiver (${\bf f}_{k}(i)={\bf 0}$)
is a convex optimization, as shown in the Appendix I, it provides
a good starting point for performing the cancellation of the
associated users by the feedforward section of the DF-CCM
receiver.

\subsection{DF Constrained Minimum Variance (DF-CMV) Receivers}

The DF-CMV receiver design resembles the DF-CCM design and
considers the following cost function :
\begin{equation}
J_{MV} = E\Big[|{\bf w}_{k}^{H}(i){\bf r}(i) -{\bf
f}_{k}^{H}(i)\hat{\bf b}(i)|^{2}\Big]
\end{equation}
subject to ${\bf C}_{k}^{H}{\bf w}_{k}(i) = {\bf h}_{k}(i)$. Given
the channel vector ${\bf h}_{k}(i)$, let us consider the
unconstrained cost function ${J}_{MV}'(i) = E\Big[|{\bf
w}_{k}^{H}(i){\bf r}(i) -{\bf f}_{k}^{H}(i)\hat{\bf
b}(i)|^{2}\Big]+ 2 \Re[ ({\bf C}_{k}^{H}{\bf w}_{k}(i)- {\bf
h}_{k}(i))^{H}\boldsymbol{\lambda}]$, where $\boldsymbol{
\lambda}$ is a vector of complex Lagrange multipliers, and
minimize ${J}_{MV}'(i)$ with respect to ${\bf w}_{k}(i)$ and ${\bf
f}_{k}(i)$ under the set of constraints ${\bf C}_{k}^{H}{\bf
w}_{k}(i) = {\bf h}_{k}(i)$. By taking the gradient terms of
$J_{MV}'(i)$ with respect to ${\bf w}_{k}(i)$ and setting them to
zero we have $\nabla J_{MV}'(i) = E[{\bf r}(i)({\bf r}^{H}(i){\bf
w}_{k}(i)-\hat{\bf b}^{H}(i){\bf f}_{k}(i))] + 2{\bf C}_{k}
\boldsymbol{\lambda} = {\bf 0}$, then rearranging the terms we
obtain $E[{\bf r}(i){\bf r}^{H}(i)] {\bf w}_{k}(i) = E[{\bf
r}(i)\hat{\bf b}^{H}(i)]{\bf f}_{k}(i) - 2{\bf C}_{k}
\boldsymbol{\lambda}$ and consequently ${\bf w}_{k}(i) = {\bf
R}^{-1}(i)[{\bf T}(i){\bf f}_{k}(i) - 2{\bf C}_{k}
\boldsymbol{\lambda}]$, where the covariance matrix is ${\bf
R}=E[{\bf r}(i){\bf r}^{H}(i)]$ and ${\bf T}(i) = E[{\bf
r}(i)\hat{\bf b}(i)]$. Using the constraint ${\bf C}_{k}^{H}{\bf
w}_{k}(i) = {\bf h}_{k}(i)$ we arrive at the expression for the
Lagrange multiplier $\boldsymbol{\lambda} =({\bf C}_{k}^{H}{\bf
R}^{-1}(i){\bf C}_{k})^{-1} ({\bf C}_{k}^{H}{\bf R}^{-1}(i){\bf
T}(i){\bf f}_{k}(i) - {\bf h}_{k}(i))/2$. By substituting
$\boldsymbol{\lambda}$ into ${\bf w}_{k}(i) = {\bf R}^{-1}(i)[{\bf
T}(i){\bf f}_{k}(i) - 2{\bf C}_{k} \boldsymbol{\lambda}]$ we
obtain the solution for the feedforward section of the DF-CMV
receiver:
\begin{equation}
{\bf w}_{k}(i) =  {\bf R}^{-1}(i) \Bigg[ {\bf T}(i){\bf
f}_{k}(i)-{\bf C}_{k} ({\bf C}_{k}^{H}{\bf R}^{-1}(i){\bf
C}_{k})^{-1} \times $$ $$({\bf C}_{k}^{H}{\bf R}^{-1}(i){\bf
T}(i){\bf f}_{k}(i) - {\bf h}_{k}(i)) \Bigg]
\end{equation}
Next, we compute the gradient terms of $J_{MV}'(i)$ with respect
to ${\bf f}_{k}(i)$ and set them to zero to get $\nabla J_{MV}'(i)
= E[\hat{\bf b}(i) ({\bf r}^{H}(i){\bf w}_{k}(i)-\hat{\bf
b}^{H}(i){\bf f}_{k}(i))] = {\bf 0}$, then rearranging the terms
we have $ E[\hat{\bf b}(i)\hat{\bf b}^{H}(i)]{\bf f}_{k}(i)
=E[\hat{\bf b}(i){\bf r}^{H}(i)] {\bf w}_{k}(i)$ and consequently
we obtain
\begin{equation}
{\bf f}_{k}(i) = {\bf B}^{-1}(i) ~\Big[{\bf T}^{H}(i){\bf
w}_{k}(i)\Big]
\end{equation}
where ${\bf B}(i) =E[\hat{\bf b}(i)\hat{\bf b}^{H}(i)]$. At this
point, the designer can avoid the inversion of ${\bf B}(i)$ by
using a judicious approximation, that is ${\bf I} \approx
E[\hat{\bf b}(i)\hat{\bf b}^{H}(i)]$ \cite{verdu}, which is
verified unless the error rate is high. Hence, the feedback
section filter can be designed as given by ${\bf f}_{k}(i) \approx
{\bf T}^{H}(i){\bf w}_{k}(i)$. It should also be noted that by
making { ${\bf f}_{k}(i)= {\bf 0}$} we arrive at the solution of
Xu and Tsatsanis in \cite{xu&tsatsanis}.

\subsection{Blind Channel Estimation}

The solutions for the CCM and CMV DF receivers assume the
knowledge of the channel parameters. However, in applications
where multipath is present these parameters are not known and thus
channel estimation is required. To blindly estimate the channel we
use the method of \cite{xu&tsatsanis,douko1}:
\begin{equation}
\hat{\bf h}_{k}(i) = \arg \min_{\bf h_{k}}~~{{\bf h}_{k}^{T} {\bf
C}^{T}_{k}{\bf R}^{-p}(i){\bf C}_{k}{\bf h}_{k}}
\end{equation}
subject to { $||\hat{\bf h}_{k}||=1$}, where $p$ is an integer and
whose solution is the eigenvector corresponding to the minimum
eigenvalue of the { $L_{p}\times L_{p}$} matrix  { $ {\bf
C}^{T}_{k}{\bf R}^{-p}(i){\bf C}_{k}$}. For the CCM receiver we
employ { ${\bf R}_{k}(i)$} in lieu of { ${\bf R}(i)$} (used for
the CMV) for channel estimation. The use of ${\bf R}_{k}(i)$
instead of ${\bf R}$ avoids the estimation of both ${\bf R}(i)$
and ${\bf R}_{k}(i)$, and shows no performance loss as verified in
our studies and explained in Appendix IV. The values of $p$ are
restricted to $1$ even though the performance of the channel
estimator and consequently of the receiver can be improved by
increasing $p$.

\section{Successive Parallel Arbitrated and Iterative DF Detection}

In this section, we present novel iterative techniques, which are
based on the recently introduced concept of parallel arbitration
\cite{barriac}, and combine them with iterative cascaded DF stages
\cite{woodward2,woodward3}. The motivation for the novel DF
structures is to mitigate the effects of error propagation often
found in P-DF structures \cite{woodward2,woodward3}, that are of
great interest for uplink scenarios due to its capability of
providing uniform performance over the users. The basic idea is to
improve the S-DF structure using parallel searches and then
combine it with an iterative technique, where the second stage
uses a P-DF system to equalize the performance of the users.

\subsection{Successive Parallel Arbitrated DF Detection}

The idea of parallel arbitration is to employ successive
interference cancellation (SIC) to rapidly converge to a local
maximum of the likelihood function and, by running parallel
branches of SIC with different orders of cancellation, one can
arrive at sufficiently different local maxima \cite{barriac}. In
order to obtain the benefits of parallel search, the candidates
should be arbitrated, yielding different estimates of a symbol.
The estimate of a symbol that has the highest likelihood is then
selected at the output.

Unlike the work of Barriac and Madhow \cite{barriac} that employed
matched filters as the starting point, we adopt blind DF receivers
as the initial condition. The concept of parallel arbitration is
thus incorporated into a DF detector structure, that applies
linear interference suppression followed by SIC and yields
improved starting points as compared to matched filters. It is
also worth noting that our approach does not require regeneration
as occurs with the original PASIC in \cite{barriac} because the
blind adaptive filters automatically compute the coefficients for
interference cancellation. A block diagram of the proposed scheme,
denoted successive parallel arbitrated decision feedback (SPA-DF),
is shown in Fig. 2.

\begin{figure}[!htb]
\begin{center}
\def\epsfsize#1#2{0.75\columnwidth}
\epsfbox{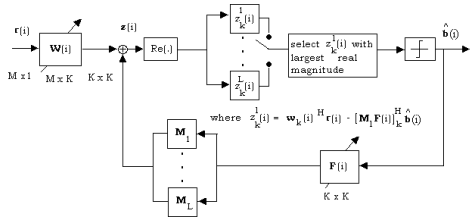} \caption{Block diagram of the proposed blind
SPA-DF receiver. }
\end{center}
\end{figure}

Following the schematics of Fig. 2, the user $k$ output of the
parallel branch $l$ ($l=1,~ \ldots,~ L$) for the SPA-DF receiver
structure is given by:
\begin{equation}
z_{k}^{l}(i) = {\bf w}^{H}_{k}(i){\bf r}(i) - [{\bf M}_{l}{\bf
F}]_{k}^{H}\hat{\bf b}(i)
\end{equation}
where the vector with initial decisions is $\hat{\bf
b}(i)=sgn[\Re({\bf W}^{H}(i){\bf r}(i))]$ and the matrices ${\bf
M}_{l}$ are permutated square identity (${\bf I}_{K}$) matrices
with dimension $K$ whose structures for an $L=4$-branch SPA-DF
scheme are given by:
\begin{equation}
{\bf M}_{1}= {\bf I}_{K},~ {\bf M}_{2}=\left[\begin{array}{cc}
{\bf 0}_{K/4,3K/4} & {\bf I}_{3K/4} \\
{\mathbf{I}}_{K/4} & {\bf 0}_{K/4,3K/4}
\end{array}\right], ~ $$ $$
{\bf M}_{3}=\left[\begin{array}{cc}
{\bf 0}_{K/2} & {\bf I}_{K/2} \\
{\bf I}_{K/2} & {\bf 0}_{K/2}
\end{array}\right],~
{\bf M}_{4}=\left[\begin{array}{ccc}
 0 & \ldots & 1 \\
\vdots & . \cdot {\Large }^{\Large .} &  \vdots \\
1 & \ldots  & 0
\end{array}\right]
\end{equation}
where ${\bf 0}_{m,n}$ denotes an $m \times n$-dimensional matrix
full of zeros and the structures of the matrices $M_{l}$
correspond to phase shifts regarding the cancellation order of the
users. Indeed, the purpose of the matrices in (18) is to change
the order of cancellation. When ${\bf M}={\bf I}$ the order of
cancellation is a simple successive cancellation (S-DF) based upon
the user powers (the same as \cite{duel,varanasi2}). Specifically,
the above matrices perform the cancellation with the following
order with respect to user powers: ${\bf M}_{1}$ with $1, \ldots,
K$; ${\bf M}_{2}$ with $K/4,K/4+1, \ldots, K,1,
\ldots,K/4-1$;${\bf M}_{3}$ with $K/2,K/2+1, \ldots,
K,1,\ldots,K/2-1$; ${\bf M}_{4}$ with $K, \ldots, 1$ (reverse
order). For more branches, additional phase shifts are applied
with respect to user cancellation ordering. It is also worth
noting that different update orders have been tried although they
did not result in performance improvements. For the proposed
SPA-DF, the number of feedback elements used and their associated
number of non-zero filter coefficients in ${\bf f}_{k}$ (where $k$
goes from the second detected user to the last one) range from $1$
to $K-1$ according to the branch $l$ and the matrix ${\bf M}_{l}$.

The final output $\hat{b}_{k}^{f}(i)$ of the SPA-DF detector
chooses the estimate of the $L$ candidates as described by:
\begin{equation}
\hat{b}_{k}^{(f)}(i) = sgn\Bigg[ \arg \max_{1 \leq l \leq L} |\Re
\Big( z_{k}^{l}(i) \Big)| \Bigg]
\end{equation}
where the selected estimate is the one with largest real
magnitude, that forms the vector of final decisions $\hat{\bf
b}_{k}^{(f)}(i)=[\hat{b}_{1}^{(f)}(i)~\ldots
~\hat{b}_{K}^{(f)}(i)]^{T}$. The number of parallel branches $L$
that yield detection candidates is a parameter that must be chosen
by the designer. Our studies and computer simulations indicate
that $L=4$ achieves most of the gains of the proposed structure
and offers a good trade-off between performance and complexity. In
terms of complexity the SPA-DF system employs the same filters,
namely ${\bf W}(i)$ and ${\bf F}(i)$, of the traditional S-DF and
requires additional arithmetic operations to compute the parallel
arbitrated candidates. As occurs with S-DF receivers, a
disadvantage of the SPA-DF detector is that it generally does not
provide uniform performance over the user population.
Specifically, in a scenario with tight power control successive
techniques tend to favor the last detected users, resulting in
non-uniform performance. To equalize the performance of the users
an iterative technique with multiple stages can be used.

\subsection{Iterative Successive Parallel Arbitrated DF Detection}

In \cite{woodward2}, Woodward {\it et al.} presented an iterative
detector with an S-DF in the first stage and P-DF or S-DF
structures, with users being demodulated in reverse order, in the
second stage. The work of \cite{woodward2} was then extended to
account for coded systems and training-based reduced-rank filters
\cite{woodward3}. Differently from \cite{woodward2,woodward3}, we
focus on blind adaptive receivers, uncoded systems and combine the
proposed SPA-DF structure with iterative detection. An iterative
receiver with hard-decision feedback is defined by the recursion:
\begin{equation}
{\bf z}^{(m+1)}(i)={\bf W}^{H}(i){\bf r}(i) - {\bf
F}^{H}(i)\hat{\bf b}^{(m)}(i)
\end{equation}
where the filters ${\bf W}$ and ${\bf F}$ can be S-DF or P-DF
structures, and $\hat{\bf b}^{m}(i)$ is the vector of tentative
decisions from the preceding iteration, where we have:
\begin{equation}
\hat{\bf b}^{(1)}(i) = sgn \Big(\Re \Big[{\bf W}^{H}(i){\bf
r}(i)\Big] \Big)
\end{equation}
\begin{equation}
\hat{\bf b}^{(m)}(i) = sgn \Big(\Re \Big[{\bf z}^{(m)}(i)\Big]
\Big), ~m>1
\end{equation}
where the number of stages $m$ depends on the application.
Additional stages can be added where the order of the users is
reversed from stage to stage.

\begin{figure}[!htb]
\begin{center}
\def\epsfsize#1#2{0.5\columnwidth}
\epsfbox{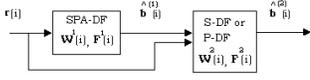} \caption{Block diagram of the two-stage DF
receiver with SPA-DF scheme in the first stage. The second stage
can employ S-DF or P-DF structures to demodulate users in reverse
order relative to the first branch of the first stage, that uses
S-DF detection. }
\end{center}
\end{figure}

To equalize the performance over the user population, we consider
the two-stage structure shown in Fig. 3. The first stage is an
SPA-DF scheme with filters ${\bf W}^{1}$ and ${\bf F}^{1}$. The
tentative decisions are passed to the second stage, which consists
of an S-DF or an P-DF detector with filters ${\bf W}^{2}$ and
${\bf F}^{2}$. The users in the second stage are demodulated
successively and in reverse order relative to the first branch of
the SPA-DF structure (a conventional S-DF). The resulting
iterative receiver system is denoted ISPAS-DF when an S-DF scheme
is deployed in the second stage, whereas for a P-DF filters in the
second stage the overall scheme is called ISPAP-DF. The output of
the second stage of the resulting scheme is expressed by:
\begin{equation}
z_{j}^{(2)}(i)=[{\bf M}{\bf W}^{2}(i)]_{j}^{H}{\bf r}(i) - [{\bf
M}{\bf F}^{2}(i)]_{j}^{H}\hat{\bf b}^{(2)}(i)
\end{equation}
where $z_{j}$ is the $j$th component of the soft output vector
${\bf z}$, ${\bf M}$ is a square permutation matrix with ones
along the reverse diagonal and zeros elsewhere (similar to ${\bf
M}_{4}$ in (18)), $[\cdot]_{j}$ denotes the $j$th column of the
argument (a matrix), and $\hat{b}_{j}^{m}(i) =
sgn[\Re(z_{j}^{m}(i))]$. Note that additional stages can be
included or the SPA-DF scheme can be used in the second stage,
even though our studies indicate that the gains in performance are
marginal. Hence, the two-stage structure is adopted for the rest
of this work. It should also be remarked that, due to the
difficulty of theoretically analyzing parallel arbitrated and
iterative schemes, our analysis in Section VI is mainly focused on
computer simulation experiments. A theoretical analysis of
iterative DF schemes constitutes an open topic which is beyond the
scope of this paper.

\section{Adaptive Algorithms}

In this section we describe stochastic gradient (SG) and recursive
least squares (RLS) algorithms for the blind estimation of the
channel, the feedforward and feedback sections of DF receivers
using the CM and MV criteria along with constrained optimization
techniques, as illustrated in Fig. 1. The CMV-based algorithms are
extensions for DF detection of the techniques proposed by Xu and
Tsatsanis in \cite{xu&tsatsanis}. The CCM-SG recursions represent
an extension of the work of \cite{xu&liu} for complex signals and
DF receivers, whereas the CCM-RLS algorithms are novel for both
linear and DF structures.

It should be emphasized that the SG solutions presented in this
section differ from those reported in a previous work
\cite{delamareel} in the sense that the blind channel estimation
is decoupled from the feedforward and feedback recursions. Indeed,
we adopt the SG blind channel estimation reported in \cite{douko2}
that has been shown to outperform the one proposed in
\cite{xu&tsatsanis}. Our studies also reveal that when the system
deals with high loads ($K$ is large) and the performance is
poorer, a decoupled SG blind channel estimator, such as
\cite{douko2}, is significantly less affected than the approach
that optimizes ${\bf w}_{k}$, ${\bf f}_{k}$ and $\hat{\bf h}_{k}$
as in \cite{delamareel} . In addition, the deployment of the SG
blind estimator of \cite{douko2} with SG CCM-based algorithms
considerably improves its performance, because blind channel
estimators that rely on the CM criterion show poor performance and
depend on other methods for initialization, as pointed out in
\cite{xu&liu}.

In terms of performance, RLS recursions have the potential to
achieve good performance independently of the spread of the
eigenvalues of the input signal autocorrelation matrix, have
faster convergence performance, show superior performance under
fast frequency selective fading channels and can cope with larger
systems \cite{10} than SG techniques.

In terms of complexity SG algorithms require a number of
operations that grows linearly with $M$ and additional users in
order to suppress MAI, ISI and estimate the channel \cite{douko2},
whereas RLS techniques have quadratic complexity implementation
for MAI, ISI suppression and channel estimation.

\subsection{Stochastic Gradient and RLS Blind Channel Estimation}

The channel estimate $\hat{\bf h}_{k}(i)$ is obtained through the
power method and the SG and RLS techniques described in
\cite{douko1}. The methods are SG and RLS adaptive version of the
blind channel estimation algorithms described in (16) and
introduced in \cite{douko2}. The SG recursion requires only
$O(L_{p}^2)$ arithmetic operations to estimate the channel,
against $O(L_{p}^{3})$ of its SVD version. For the RLS version,
the SVD on the $L_{p}\times L_{p}$ matrix ${\bf C}_{k}^{H}{\bf
R}^{-1}(i){\bf C}_{k}$, as stated in (16) and that requires
$O(L_{p}^3)$, is avoided and replaced by a single matrix-vector
multiplication, resulting in the reduction of the corresponding
computational complexity on one order of magnitude and no
performance loss. For the CCM-RLS algorithms, ${\bf R}_{k}$ can be
employed instead of ${\bf R}$ (used for the CMV) for channel
estimation to avoid the estimation of both ${\bf R}$ and ${\bf
R}_{k}$. The use of ${\bf R}_{k}$ instead of ${\bf R}$ shows no
performance loss as verified in our studies and is explained in
Appendix IV.

\subsection{Constrained Constant Modulus Stochastic
Gradient (CCM-SG) Algorithm }

An SG solution to (10) and (11) can be devised by using
instantaneous estimates and taking the gradient terms with respect
to ${\bf w}_{k}(i)$ and ${\bf f}_{k}(i)$ which should adaptively
minimize $J_{CM}$ with respect to ${\bf w}_{k}(i)$ and ${\bf
f}_{k}(i)$. The recursions of \cite{douko2} are used to obtain
channel estimates. If we consider the set of constraints ${\bf
C}_{k}^{H}{\bf w}_{k}(i) = \hat{\bf h}_{k}(i)$, we arrive at the
update equations for the estimation of ${\bf w}_{k}(i)$ and ${\bf
f}_{k}(i)$:
\begin{equation}
{\bf w}_{k}(i+1) = {\bf P}_{k}({\bf w}_{k}(i) -
\mu_{w}e_{k}(i)z_{k}^{*}(i){\bf r}(i))+ \nu~{\bf C}_{k} ({\bf
C}_{k}^{H}{\bf C}_{k})^{-1}\hat{\bf h}_{k}(i)
\end{equation}
\begin{equation}
{\bf f}_{k}(i+1) = {\bf f}_{k}(i) -
\mu_{f}e_{k}(i)z_{k}^{*}(i)\hat{\bf b}(i)
\end{equation}
where $z_{k}(i) = {\bf w}_{k}^{H}(i){\bf r}(i)-{\bf
f}_{k}^{H}(i)\hat{\bf b}(i)$, $e_{k}(i) = (|z_{k}(i)|^{2}-1)$ and
${\bf P}_{k} = {\bf I} -{\bf C}_{k}({\bf C}_{k}^{H}{\bf
C}_{k})^{-1}{\bf C}_{k}^{H}$ is a matrix that projects the
receiver's parameters onto another hyperplane in order to ensure
the constraints.

It is worth noting that, for stability and to facilitate tuning of
parameters, it is useful to employ normalized step sizes when
operating in a changing environment. A normalized version of this
algorithm can be devised by substituting (24) and (25) into the CM
cost function, differentiating the cost function with respect to
$\mu_{w}$ and $\mu_{f}$, setting it to zero and solving the new
equations, as detailed in Appendix II. Hence, the normalized
CCM-SG algorithm proposed here adopts variable step size
mechanisms described by { $\mu_{w} =
\frac{\mu_{0_{w}}(|z_{k}(i)|-\mu_{f}|z_{k}(i)|e_{k}(i)\hat{\bf
b}^{H}(i)\hat{\bf b}(i)+1)} {|z_{k}(i)|e_{k}(i){\bf r}^{H}(i){\bf
P}{\bf r}(i)}$} and { $\mu_{f} =
\frac{\mu_{0_{f}}(|z_{k}(i)|-\mu_{w}|z_{k}(i)|e_{k}(i){\bf
r}^{H}(i){\bf P}{\bf r}(i)+1)  } {|z_{k}(i)|e_{k}(i) \hat{\bf
b}^{H}(i)\hat{\bf b}(i)}$} where $\mu_{0_{w}}$ and $\mu_{0_{f}}$
are the convergence factors for ${\bf w}_{k}$ and ${\bf f}_{k}$,
respectively.

\subsection{Constrained Minimum Variance Stochastic Gradient (CMV-SG) Algorithm}

An SG solution to (13) and (14) can be developed in an analogous
form to the previous section by taking the gradient terms with
respect to ${\bf w}_{k}(i)$ and ${\bf f}_{k}(i)$. The recursions
in \cite{douko2} are used again to obtain channel estimates. The
update rules for the estimation of the parameters of the
feedforward and feedback sections of the DF receiver are:
\begin{equation}
{\bf w}_{k}(i+1) = {\bf P}_{k}({\bf w}_{k}(i) -
\mu_{w}z_{k}^{*}(i){\bf r}(i))+{\bf C}_{k} ({\bf C}_{k}^{H}{\bf
C}_{k})^{-1}\hat{\bf h}_{k}(i)
\end{equation}
\begin{equation}
{\bf f}_{k}(i+1) = {\bf f}_{k}(i) - \mu_{f}z_{k}^{*}(i)\hat{\bf
b}(i)\end{equation}

A normalized version of this algorithm can also be obtained by
substituting (26) and (27) into the MV cost function,
differentiating it with respect to $\mu_{w}$ and $\mu_{f}$,
setting it to zero and solving the new equations, as described in
Appendix III. Hence, $\mu_{w} = \frac{\mu_{0_{w}}(1- \mu_{f}
\hat{\bf b}^{H}(i)\hat{\bf b}(i))} {{\bf r}^{H}(i){\bf P}{\bf
r}(i)}$ and $\mu_{f} = \frac{\mu_{0_{f}}(1 - \mu_{w} {\bf
r}^{H}(i){\bf P}{\bf r}(i)) } {\hat{\bf b}^{H}(i)\hat{\bf b}(i)}$.

\subsection{Constrained Constant Modulus RLS (CCM-RLS) Algorithm}

Given the expressions for the feedforward ({ ${\bf w}_{k}$}) and
feedback ({ ${\bf f}_{k}$}) sections in (11) and (12) of the blind
DF receiver, we need to estimate { ${\bf R}^{-1}_{k}(i)$, ${\bf
I}^{-1}_{k}(i)$ } and { $({\bf C}_{k}^{T}{\bf R}^{-1}_{k}(i){\bf
C}_{k})^{-1}$} recursively to reduce the computational complexity
required to invert these matrices. Using the matrix inversion
lemma and Kalman RLS recursions \cite{10} we have:
\begin{equation}
{\bf G}_{k}(i) =  \frac{\alpha^{-1}\hat{\bf R}_{k}^{-1}(i-1)
z_{k}^{*}(i){\bf r}(i)} {1+ \alpha^{-1} {\bf r}^{H}(i) z_{k}(i)
\hat{\bf R}_{k}^{-1}(i-1)z_{k}^{*}(i) {\bf r}(i)}
\end{equation}
\begin{equation}
\hat{\bf R}_{k}^{-1}(i) = \alpha^{-1} \hat{\bf R}_{k}^{-1}(i-1)
-\alpha^{-1}{\bf G}_{k}(i) z_{k}(i) {\bf r}^{H}(i) \hat{\bf
R}_{k}^{-1}(i-1)
\end{equation}
and
\begin{equation}
{\bf V}(i) =  \frac{\alpha^{-1} \hat{\bf I}^{-1}_{k}(i-1)
z_{k}^{*}(i) \hat{\bf b}(i)} {1+ \alpha^{-1} \hat{\bf b}^{H}(i)
z_{k}(i) \hat{\bf I}^{-1}_{k}(i-1)z_{k}^{*}(i) \hat{\bf b}(i)}
\end{equation}
\begin{equation}
\hat{\bf I}^{-1}_{k}(i) = \alpha^{-1} \hat{\bf I}^{-1}_{k}(i-1)
-\alpha^{-1}{\bf V}(i) z_{k}(i) \hat{\bf b}^{H}(i) \hat{\bf
I}^{-1}_{k}(i-1)
\end{equation}
where $0<\alpha < 1$ is the forgetting factor. The algorithm can
be initialized with { $\hat{\bf R}_{k}^{-1}(0)=\delta {\bf I}$}
and { $\hat{\bf I}^{-1}_{k}(0)=\delta {\bf I}$} where $\delta$ is
a scalar to ensure numerical stability. Once { ${\bf
R}_{k}^{-1}(i)$} is updated, we employ another recursion to
estimate $({\bf C}_{k}^{H}{\bf R}^{-1}_{k}(i){\bf C}_{k})^{-1}$ as
described by:
\begin{equation}
\boldsymbol{\Gamma}^{-1}_{k}(i) =
\frac{\boldsymbol{\Gamma}^{-1}_{k}(i-1)}{1-\alpha}-
\frac{\boldsymbol{\Gamma}^{-1}_{k}(i-1)
\boldsymbol{\gamma}_{k}(i)\boldsymbol{\gamma}^{H}_{k}(i)\boldsymbol{\Gamma}^{-1}_{k}(i-1)}
{\frac{(1-\alpha)^2}{\alpha}+(1-\alpha)\boldsymbol{\gamma}^{H}_{k}(i)\boldsymbol{\Gamma}^{-1}_{k}(i)\boldsymbol{\gamma}_{k}(i)}
 \end{equation} where
$\boldsymbol{\Gamma}_{k}(i)$ is an estimate of $({\bf
C}_{k}^{H}{\bf R}^{-1}_{k}(i){\bf C}_{k})$ and
$\boldsymbol{\gamma}_{k}(i)={\bf C}_{k}^{H}{\bf r}(i)z_{k}(i)$.
The RLS channel estimation procedure described in \cite{douko2}
with $\boldsymbol{\Gamma}_{k}$ in lieu of
$\boldsymbol{\Theta}_{k}$ is employed for estimating ${\bf
h}_{k}$, saving computational resources and resulting in no
performance loss for channel estimation. Finally, we construct the
DF-CCM receiver as described by:
\begin{equation}
\hat{\bf w}_{k}(i) = \hat{\bf R}_{k}^{-1}(i)\Big[\hat{\bf
d}_{k}(i) + \hat{\bf T}_{k}(i)\hat{\bf f}_{k}(i) - {\bf C}_{k}
\hat{\boldsymbol{\Gamma}}^{-1}(i) \times $$ $$ \Big({\bf
C}_{k}^{H}\hat{\bf R}_{k}^{-1}(i)\hat{\bf T}_{k}(i){\bf f}_{k}(i)
+ {\bf C}_{k}^{H}\hat{\bf R}_{k}^{-1}(i)\hat{\bf d}_{k}(i) -
\nu~\hat{\bf h}_{k}(i)\Big)\Big]
\end{equation}
\begin{equation}
\hat{\bf f}_{k}(i) = {\bf I}_{k}^{-1}(i)\Big[\hat{\bf
T}_{k}^{H}(i)\hat{\bf w}_{k}(i) - \hat{\bf v}_{k}(i)\Big]
\end{equation}
where  ${\bf d}_{k}(i)$ is estimated by $\hat{\bf
d}_{k}(i+1)=\alpha \hat{\bf d}_{k}(i)+(1-\alpha)z_{k}^{*}(i){\bf
r}(i)$,  $\hat{\bf T}_{k}(i+1)=\alpha\hat{\bf T}_{k}(i)
+(1-\alpha)\hat{\bf b}_{k}(i){\bf r}^{H}(i)|z_{k}(i)|^{2}$ and
$\hat{\bf v}_{k}(i+1)=\alpha\hat{\bf
v}_{k}(i)+(1-\alpha)z_{k}^{*}(i)\hat{\bf b}(i)$. In terms of
computational complexity, the CCM-RLS algorithm requires
$O(M^{2})$ (feedforward section) and $O(K^{2})$ (feedback section)
to suppress MAI and ISI and $O(L_{p}^{2})$ to estimate the
channel, against $O(M^{3})$, $O(K^{3})$ and $O(L_{p}^{3})$
required by (11), (12) and (16), respectively.

\subsection{Constrained Minimum Variance RLS (CMV-RLS)  Algorithm}

Similarly to the CCM-RLS, the expressions for the DF-CMV receiver
given in (14) and (15) are employed, and the matrices ${\bf
R}^{-1}(i)$, ${\bf B}^{-1}(i)$ and { $({\bf C}_{k}^{T}{\bf
R}^{-1}(i){\bf C}_{k})^{-1}$ are recursively estimated with the
aid of the matrix inversion lemma in order to reduce the
computational complexity as given by :
\begin{equation}
{\bf G}(i) =  \frac{\alpha^{-1}\hat{\bf R}^{-1}(i-1){\bf r}(i)}
{1+ \alpha^{-1} {\bf r}^{H}(i) \hat{\bf R}^{-1}(i-1){\bf r}(i)}
\end{equation}
\begin{equation}
\hat{\bf R}^{-1}(i) = \alpha^{-1} \hat{\bf R}^{-1}(i-1)
-\alpha^{-1}{\bf G}(i) {\bf r}^{T}(i) \hat{\bf R}^{-1}(i-1)
\end{equation}
and
\begin{equation}
{\bf Q}(i) =  \frac{\alpha^{-1} \hat{\bf B}^{-1}(i-1) \hat{\bf
b}(i)} {1+ \alpha^{-1} \hat{\bf b}^{H}(i) \hat{\bf
B}^{-1}_{k}(i-1) \hat{\bf b}(i)}
\end{equation}
\begin{equation}
\hat{\bf B}^{-1}(i) = \alpha^{-1} \hat{\bf B}^{-1}(i-1)
-\alpha^{-1}{\bf Q}(i) \hat{\bf b}^{H}(i) \hat{\bf B}^{-1}(i-1)
\end{equation}
where $0<\alpha < 1$ is the forgetting factor. The algorithm can
be initialized with { $\hat{\bf R}^{-1}(0)=\delta {\bf I}$} and {
${\bf B}^{-1}(0)=\delta {\bf I}$} where $\delta$ is a positive
constant. Once $\hat{\bf R}^{-1}(i)$ is updated, we employ another
recursion to estimate $({\bf C}_{k}^{H}\hat{\bf R}^{-1}(i){\bf
C}_{k})^{-1}$ as described by:
\begin{equation}
\boldsymbol{\Theta}^{-1}_{k}(i) = \Bigg[
\frac{\boldsymbol{\Theta}^{-1}_{k}(i-1)}{1-\alpha}-
\frac{\boldsymbol{\Theta}^{-1}_{k}(i-1)
\boldsymbol{\theta}_{k}(i)\boldsymbol{\theta}^{H}_{k}(i)\boldsymbol{\Theta}^{-1}_{k}(i-1)}
{\frac{(1-\alpha)^2}{\alpha}+(1-\alpha)\boldsymbol{\theta}^{H}_{k}(i)\boldsymbol{\Theta}^{-1}_{k}(i)\boldsymbol{\theta}_{k}(i)}\Bigg]
 \end{equation} where
$\boldsymbol{\Theta}_{k}(i)$ is an estimate of $({\bf
C}_{k}^{H}{\bf R}^{-1}(i){\bf C}_{k})$ and
$\boldsymbol{\theta}_{k}(i)={\bf C}_{k}^{H}{\bf r}(i)$. For
estimating the channel ${\bf h}_{k}$, the RLS algorithm described
in \cite{douko2} is employed. Finally, we construct the DF-CMV
receiver as given by:
\begin{equation}
\hat{\bf w}_{k}(i) =  {\bf R}^{-1}(i) \Big[ \hat{\bf T}(i)\hat{\bf
f}_{k}(i)-{\bf C}_{k} \boldsymbol{\Theta}^{-1}_{k}(i) \times $$ $$
\Big({\bf C}_{k}^{H}\hat{\bf R}^{-1}(i)\hat{\bf T}(i)\hat{\bf
f}_{k}(i) - \hat{\bf h}_{k}(i)\Big) \Big]
\end{equation}
\begin{equation}
\hat{\bf f}_{k}(i)= \hat{\bf B}^{-1}(i) ~\Big[\hat{\bf
T}^{H}(i)\hat{\bf w}_{k}(i)\Big]
\end{equation}
where $\hat{\bf T}(i+1)=\alpha\hat{\bf T}(i) +(1-\alpha)\hat{\bf
b}_{k}(i){\bf r}^{H}(i)$. It should be remarked that the
approximation on $\hat{\bf B}$, that is ${\bf I} \approx
E[\hat{\bf b}\hat{\bf b}^{H}]$, can be used when the error rate is
low in order to avoid the matrix computations in (37) and (38).
Otherwise, in the case of moderate to high error rate, it is
preferable to employ (37) and (38) in order to guarantee adequate
performance of the algorithm.

\section{Simulations}

In this section, we evaluate the performance of the iterative
arbitrated DF structures introduced in Section IV and the blind
adaptive algorithms presented in Section V. Due to the extreme
difficulty of theoretically analyzing such scheme, we adopt a
simulations approach and conduct several experiments in order to
verify the effectiveness of the proposed techniques. In
particular, we have carried out experiments under stationary and
non-stationary scenarios to assess convergence performance in
terms of the the bit error rate (BER) of the proposed structure
and algorithms and compared them with other recently reported
algorithms and structures. Moreover, BER performance of the
receivers employing the analyzed techniques is assessed for
different loads, channel paths ($L_{p}$) and profiles, and fading
rates. The DS-CDMA system employs Gold sequences of length $N=31$.

Because we focus on uplink scenarios, users experiment different
channels. All channels assume that $L_{p}=6$ as an upper bound. We
use three-path channels with relative powers $p_{k,l}$ given by
$0$, $-3$ and $-6$ dB, where in each run and for each user the
second path delay ($\tau_{2}$) is given by a discrete uniform
random variable (r. v.) between $1$ and $4$ chips and the third
path delay is taken from a discrete uniform r. v. between $1$ and
$5-\tau_{2}$ chips. It is also assumed here that the channels
experienced by different users are statistically independent and
identically distributed. The sequence of channel coefficients for
each user $k$ ($k=1,\ldots,K$) is $h_{k,l}(i)=p_{k,l}
\alpha_{k,l}(i)$ ($l=0,1,2,~\ldots$), where $\alpha_{k,l}(i)$, is
a complex Gaussian random sequence obtained by passing complex
white Gaussian noise through a filter with approximate transfer
function $c/\sqrt{1-(f/f_{d})^{2}}$ where $c$ is a normalization
constant, $f_{d}=v/\lambda$ is the maximum Doppler shift,
$\lambda$ is the wavelength of the carrier frequency, and $v$ is
the speed of the mobile \cite{11}. This procedure corresponds to
the generation of independent sequences of correlated unit power
complex Gaussian random variables ($E[|\alpha ^2_{k,l}(i)| ]=1$)
with the path weights $p_{k,l}$ normalized so that
$\sum_{l=1}^{L_{p}} p_{k,l}^{2}=1$. The phase ambiguity derived
from the blind channel estimation method in \cite{douko2} is
eliminated in our simulations by using the phase of { ${\bf
g}(0)$} as a reference to remove the ambiguity and for fading
channels we assume ideal phase tracking and express the results in
terms of the normalized Doppler frequency $f_{d}T$
(cycles/symbol). Alternatively, differential modulation can be
used to account for the phase rotations.

In the following experiments, it is indicated the type of adaptive
algorithms used (SG or RLS), the design criterion (CCM or CMV) and
the structure (linear (L) or decision feedback (DF)). For linear
receivers (L) and their algorithms we make ${\bf f}_{k}(i)={\bf
0}$ and $\mu_{f}=0$. Amongst the analyzed DF structures, we
consider:

\begin{itemize}
\item{S-DF: the successive DF detector of \cite{duel,varanasi2}.}

\item{P-DF: the parallel DF detector of \cite{woodward1,woodward2}.}

\item{ISS-DF: the iterative system of Woodward {\it et al.} \cite{woodward2}
with S-DF in the first and second stages.}
\item{ISP-DF: the iterative system of Woodward {\it et al.} \cite{woodward2}
with S-DF in the first stage and P-DF in the second stage.}
\item{SPA-DF: the proposed successive parallel arbitrated receiver.}
\item{ISPAS-DF: the proposed iterative detector with the novel SPA-DF
in the first stage and the S-DF in the second stage.}
\item{ISPAP-DF: the proposed iterative receiver with the SPA-DF in the
first stage and the P-DF in the second stage.}

\end{itemize}

For the CCM based algorithms, we employ $\nu=1$ in order to ensure
convexity. The experiments are averaged over $200$ experiments and
the parameters of the algorithms are optimized for each scenario.
We stress that the results are shown in Figs. 4 to 8 in terms of
the average BER \cite{1} and average BER amongst the $K$ users in
the system, except for Figs. 9 and 10, where the individual BER
performance of each user is shown.

\subsection{BER Convergence Performance}

In what follows, we assess the average BER convergence performance
of the analyzed adaptive DF receiver techniques and algorithms.
The BER convergence performance of the receivers is shown for SG
and RLS algorithms, in Figs. 4 and 5, respectively. We consider a
non-stationary scenario, where the system starts with $K=8$ users
and at time $i=800$, $4$ additional users enter the system,
totalling $K=12$ users, and the blind adaptive algorithms are
subject to new interferers/users in the environment. For the sake
of comparison, we also include the curves for supervised NLMS and
RLS \cite{10} adaptive algorithms, which are trained with $200$
symbols provided by a pilot channel (at $i=1,\ldots,200$ and
$i=801,\ldots,1000$) and then switch to decision-directed mode. It
is assumed that the system has an ideal power control and signals
of the different users reach the base station with the same
average ${E}_{b}/N_{0}$. Note that given the performance of
current power control algorithms, ideal power control is not far
from a realistic situation.

\begin{figure}[!htb]
\begin{center}
\def\epsfsize#1#2{0.9\columnwidth}
\epsfbox{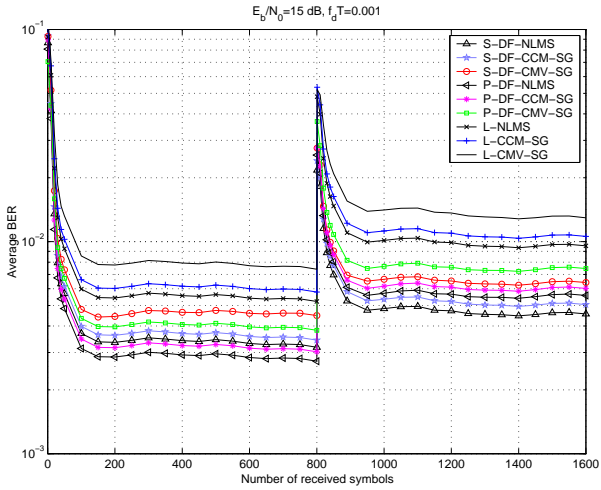} \caption{BER convergence performance of SG
algorithms.}
\end{center}
\end{figure}

The algorithms for DF receivers are initialized with a feedforward
filter ${\bf w}_{k}$ equal to the signature sequence and a
feedback filter ${\bf f}_{k}$ with zeros and they gradually adapt
in order to cancel the interference. Note that they do not lock to
an undesired user because of the blind channel estimation that
allows the receiver to use the effective signature sequence. The
results indicate that the CCM design criterion is superior to the
CMV approach, for both SG and RLS algorithms. Another conclusion
from the curves in Figs. 4 and 5 is that CCM-based blind
algorithms achieve a performance very close to the trained
algorithms, leading to significant savings in spectral efficiency.
Regarding the structures of the receivers, we note that DF
receivers are significantly better than linear detectors. In fact,
we attack the problem of the receivers presented in
\cite{xu&tsatsanis,xu&liu} that operate well in lightly loaded
systems, but do not perform well in moderate and heavily loaded
situations, by cancelling the interferers with the DF section. In
particular, P-DF schemes outperform S-DF in low BER situations,
whereas for moderate to higher BER levels S-DF systems are less
affected by error propagation.

\begin{figure}[!htb]
\begin{center}
\def\epsfsize#1#2{0.9\columnwidth}
\epsfbox{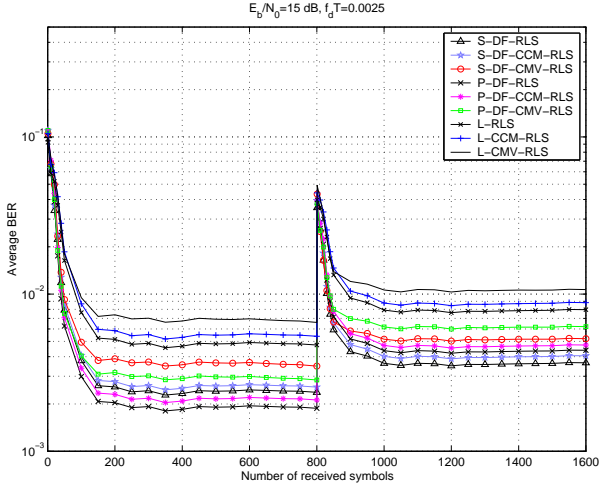} \caption{BER convergence performance of RLS
algorithms.}
\end{center}
\end{figure}

Another important conclusion from our studies is that RLS
algorithms can deal with faster fading rates and effectively
accommodate more users in the system at the cost of a quadratic
complexity, whereas SG techniques cannot deal with large systems
or very high load ($K/N$ close to $1$). Because the scenario in
the experiments assumed ideal power control, the SG algorithms
present a good convergence performance although for scenarios
without power control (near-far situations) the performance of
these algorithms is subject to the eigenvalue spread of the
covariance matrix of the received vector ${\bf r}(i)$.
Specifically, when the eigenvalue spread of the covariance matrix
of the received vector ${\bf r}(i)$ is large SG algorithms perform
poorly, whereas the rate of convergence of RLS algorithms is
invariant to such situation \cite{10}. Hence, for large systems or
those that do not have good power control RLS recursions are the
most appropriate.

Let us now consider the proposed SPA-DF and the combined iterative
DF system, namely ISPAP-DF and ISPAS-DF. Simulation experiments
with RLS algorithms were conducted to determine how many
arbitrated branches should be used and to account for the impact
of additional branches upon performance. We designed the novel DF
receivers with $L=2,4,8$ parallel branches and compared their BER
performance with the existing ISS-DF and ISP-DF structures, as
depicted in Fig. 6. The results show that the novel SPA-DF,
ISPAP-DF and ISPAS-DF significantly outperform the ISS-DF and
ISP-DF structures, and their performances improve as the number of
parallel branches increase. In this regard, we also notice that
the gains of performance obtained through additional branches
decrease as additional branches are added, resulting in marginal
improvements for more than $L=4$ branches. For this reason, we
adopt $L=4$ for the remaining experiments because it presents a
very attractive trade-off between performance and complexity.
Another conclusion from the curves in Fig. 6 is that the proposed
SPA-DF, ISPAP-DF and ISPAS-DF receiver techniques obtain
substantial gains in performance over existing iterative DF
techniques, namely, the ISP-DF and the ISS-DF of \cite{woodward2}.

\begin{figure}[!htb]
\begin{center}
\def\epsfsize#1#2{0.9\columnwidth}
\epsfbox{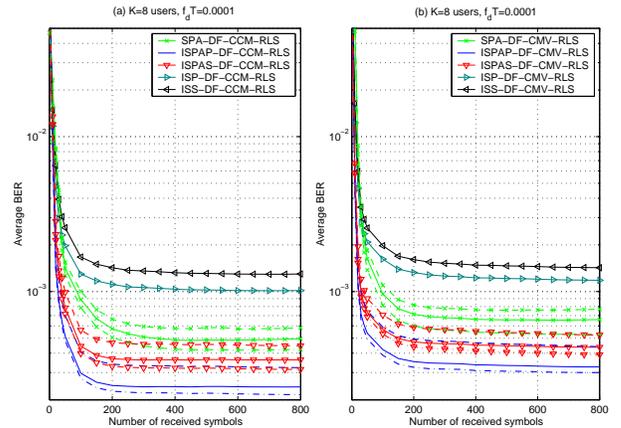} \caption{BER convergence performance of RLS
algorithms with iterative receivers for different numbers of
parallel branches $L$ at $E_{b}/N_{0}=15$ dB in a slow fading
environment (a) CCM (b) CMV (L=2 - dashed line, L=4 - solid line
and L=8 - dash-dotted line) .}
\end{center}
\end{figure}

\subsection{BER Performance versus $E_{b}/N_{0}$, K and User Index}

In this part, the BER performance of the different receiver
techniques is further investigated and the receivers process
$2000$ symbols to obtain the curves. In particular, the average
BER performance of the receivers versus $E_{b}/N_{0}$ and number
of users (K) is depicted in Figs. 7 and 8, whereas the individual
BER performance versus the user indices is shown in Figs. 9 and
10.

\begin{figure}[!htb]
\begin{center}
\def\epsfsize#1#2{0.9\columnwidth}
\epsfbox{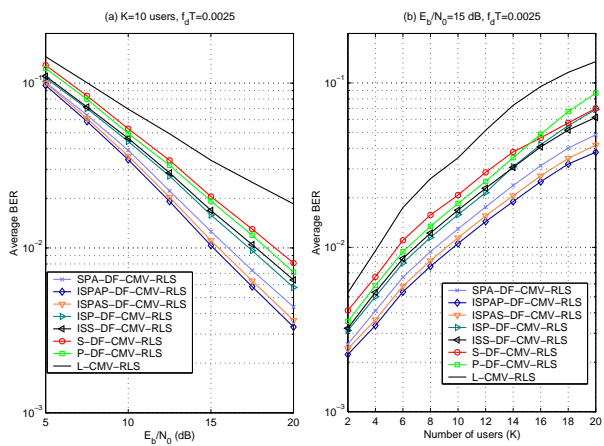} \caption{Performance of CMV-RLS algorithms in a
dynamic environment in terms of BER versus (a) $E_{b}/N_{0}$ with
$K=10$ users and (b) number of users (K) at $E_{b}/N_{0}=15$ dB.}
\end{center}
\end{figure}

\begin{figure}[!htb]
\begin{center}
\def\epsfsize#1#2{0.9\columnwidth}
\epsfbox{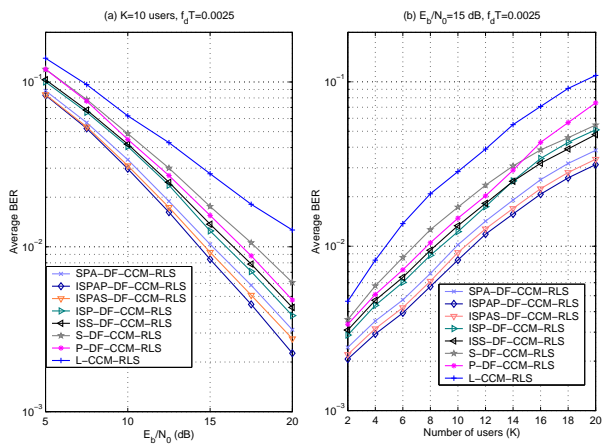} \caption{Performance of CCM-RLS algorithms in a
dynamic environment in terms of BER versus (a) $E_{b}/N_{0}$ with
$K=10$ users and (b) number of users (K) at $E_{b}/N_{0}=15$ dB.}
\end{center}
\end{figure}

A comparison of the CMV and the CCM design criteria with RLS
algorithms is carried out in experiments whose results are shown
in Figs. 7 and 8. The curves reveal that DF detectors are
significantly superior to linear receivers and that the CCM-RLS
algorithm outperforms the CMV-RLS techniques in all situations.
With respect to the performance, the best results are obtained
with the ISPAP-DF receiver structure, that can save up to $2.5$ dB
for the same BER as compared to the iterative receivers of
\cite{woodward2} (ISP-DF and ISS-DF). In comparison with linear
receivers, the proposed ISPAP-DF system obtains savings of up to
$7$ dB for the BER performance. In general, the curves in Figs. 7
and 8 reveal that the novel iterative arbitrated DF schemes,
namely the SPA-DF, ISPAP-DF and ISPAS-DF, can offer considerable
gains as compared to existing DF and linear receivers and support
systems with higher loads, through mitigation of the effects of
error propagation.

\begin{figure}[!htb]
\begin{center}
\def\epsfsize#1#2{0.9\columnwidth}
\epsfbox{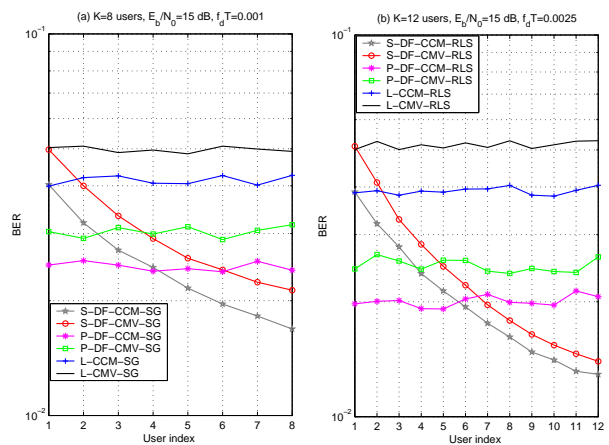} \caption{Performance of the receivers in a
fading environment in terms of individual BER versus user index
for (a) SG and (b) RLS algorithms.}
\end{center}
\end{figure}

\begin{figure}[!htb]
\begin{center}
\def\epsfsize#1#2{0.9\columnwidth}
\epsfbox{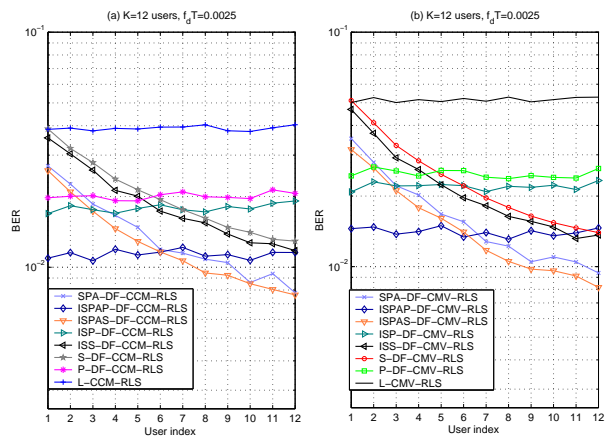} \caption{Performance of the receivers in terms
of individual BER versus user index for (a) CCM-RLS and (b)
CMV-RLS algorithms.}
\end{center}
\end{figure}

The last two scenarios, shown in Figs. 9 and 10, consider the
individual BER performance of the users. From the results, we
observe that a disadvantage of S-DF relative to P-DF is that it
does not provide uniform performance over the user population. We
also notice that for the S-DF receivers, user $1$ achieves the
same performance of their linear receivers counterparts, and as
the successive cancellation is performed users with higher indices
benefit from the interference cancellation. The same non-uniform
performance is verified for the proposed SPA-DF, the existing
ISS-DF and the novel ISPAS-DF. Conversely, the new ISPAP-DF, the
existing P-DF and the existing ISP-DF provide uniform performance
over the users which is an important goal for the uplink of
DS-CDMA systems. In particular, the novel ISPAP-DF detector
achieves the best performance of the analyzed structures and is
significantly superior to the ISP-DF and to the P-DF, that suffers
from error propagation.

\section{Conclusions}

Blind adaptive SG and RLS type algorithms based on the CMV and CCM
performance criteria were developed for estimating the parameters
of DF receivers in uplink scenarios with multipath. The CCM-based
blind algorithms have shown a performance that is very close to
that of trained algorithms without the need for pilot channels. A
novel SPA-DF structure was presented and combined with iterative
techniques for use with cascaded DF stages, resulting in new
iterative DF schemes, the ISPAS-DF and the ISPAP-DF, that can
offer substantial gains in performance over existing linear and DF
detectors and mitigate more effectively the deleterious effects of
error propagation. In particular, the proposed ISPAP-DF structure
has achieved the best performance amongst all analyzed receivers
and is able to provide uniform performance over the user
population.

\begin{appendix}

\section{Convergence Properties}

In what follows, an analysis of the CCM method and its convergence
properties is carried out for the linear receiver case (${\bf
f}_{k}={\bf 0}$), extending previous results on its convexity for
both complex and multipath signals. We believe that it provides a
good starting point (better than the CMV design) for performing
the cancellation of the associated users by the feedforward
section of the DF-CCM receiver.

Let us express the cost function $J_{CM} = E\Big[(|{\bf
w}_{k}^{H}{\bf r}|^{2}-1)^{2}\Big]$ as $J_{CM} =
(E[|z_{k}|^{4}]-2E[|z_{k}|^{2}]+1)$, drop the time index (i) for
simplicity, assume a stationary scenario and that $b_{k}$,
k=1,\ldots,K are statistically independent i.i.d complex random
variables with zero mean and unit variance, $b_{k}$ and ${\bf n}$
are statistically independent. Let us also define ${\bf x}=
\sum_{k=1}^{K}A_{k} {b}_{k}\tilde{\bf s}_{k}$, ${\bf C}_{k}{\bf
h}_{k} = \tilde{\bf s}_{k}$, ${\bf Q}=E[{\bf x}{\bf x}^{H}]$,
${\bf P}=E[\boldsymbol{\eta}\boldsymbol{\eta}^{H}]$, ${\bf R}={\bf
Q}+ {\bf P} + \sigma^{2}{\bf I}$ and alternatively express the
received vector by ${\bf r}(i) = {\bf x}(i) + \boldsymbol{\eta}(i)
+ {\bf n}(i)$, where $\boldsymbol{\eta}(i)$ is the ISI.
Considering user $1$ as the desired one we let ${\bf w}_{1}={\bf
w}$ and define $u_{k}=A_{k}^{*}\tilde{\bf s}_{k}^{H}{\bf w}$,
${\bf u}={\bf A}^{H}\tilde{\bf S}^{H}{\bf w}=[u_{1} \ldots
u_{K}]^{T}$, where $\tilde{\bf S}=[\tilde{\bf s}_{1} \ldots
\tilde{\bf s}_{K}]$, ${\bf A}=diag(A_{1} \ldots A_{k})$ and ${\bf
b} =[b_{1} \ldots b_{K}]^{T}$. Using the constraint ${\bf
C}_{1}^{H}{\bf w}= \nu \hat{\bf h}_{1}$ we have for the desired
user the condition $u_{1}=(A_{1}^{*}\tilde{\bf s}_{1}^{H}){\bf w}
= A_{1}^{*}{\bf h}{\bf C}_{1}^{H}{\bf w}= \nu A_{1}^{*}{\bf
h}_{1}^{H}\hat{\bf h}_{1} $. In the absence of noise and
neglecting ISI, the (user $1$) cost function can be expressed as $
J_{CM}({\bf w})= E[({\bf u}^{H}{\bf b}{\bf b}^{H}{\bf u})^{2}] -
2E[({\bf u}^{H}{\bf b}{\bf b}^{H}{\bf u})] +1 =
8(\sum_{k=1}^{K}u_{k}u_{k}^{*})^{2} - 4
\sum_{k=1}^{K}(u_{k}u_{k}^{*})^{2} - 4
\sum_{k=1}^{K}u_{k}u_{k}^{*} + 1 = 8 ( D + \sum_{k=2}^{K}
u_{k}u_{k}^{*})^{2} - 4D^{2} - 4\sum_{k=2}^{K}
(u_{k}u_{k}^{*})^{2} - 4D -4 \sum_{k=2}^{K} (u_{k}u_{k}^{*}) + 1$,
where $D = u_{1}u_{1}^{*}=\nu^{2} |A_{1}|^{2}|\hat{\bf
h}_{1}^{H}{\bf h}_{1}|^{2}$. To examine the convergence properties
of the optimization problem in (10), we proceed similarly to
\cite{kwak}. Under the constraint ${\bf C}_{1}^{H}{\bf w}= \nu
\hat{\bf h}_{1}$, we have:
\begin{equation}
\begin{split}
J_{CM}({\bf w})=\tilde{J}_{CM}(\bar{\bf u})= & 8 ( D + \bar{\bf
u}^{H}\bar{\bf u})^{2} - 4D^{2} - 4
\sum_{k=2}^{K}(u_{k}u_{k}^{*})^{2} \\ & - 4D -4 (\bar{\bf
u}^{H}\bar{\bf u}) + 1
\end{split}
\end{equation}
where $\bar{\bf u}=[u_{2},\ldots, u_{K}]^{T}={\bf B}{\bf w}$,
${\bf B}= {\bf A}'^{H}\tilde{\bf S}'^{H}$, $\tilde{\bf
S}'=[\tilde{\bf s}_{2} \ldots \tilde{\bf s}_{K}]$ and ${\bf
A}'=diag(A_{2} \ldots A_{K})$. To evaluate the convexity of
$\tilde{J}_{CM}(.)$, we compute its Hessian (${\bf H}$) using the
rule ${\bf H} = \frac{\partial }{\partial \bar{\bf u}^{H}}
\frac{\partial (\tilde{J}_{CM}(\bar{\bf u}))}{\partial \bar{\bf
u}}$ that yields:
\begin{equation}
{\bf H} = \Big[ 16 (D -1/4){\bf I} + 16 \bar{\bf u}^{H}\bar{\bf
u}{\bf I} + 16 \bar{\bf u}\bar{\bf u}^{H} - 16diag(|u_{2}|^{2}
\ldots |u_{K}|^{2}) \Big]
\end{equation}
Specifically, ${\bf H}$ is positive definite if ${\bf a}^{H}{\bf
H}{\bf a}> 0$ for all nonzero ${\bf a} \in
\boldsymbol{C}^{K-1\times K-1}$ \cite{10}. The second, third and
fourth terms of (46) yield the positive definite matrix
$16\Big(\bar{\bf u}\bar{\bf u}^{H} +
diag(\sum_{k=3}^{K}|u_{k}|^{2}~\sum_{k=2,k\neq 3}^{K}|u_{k}|^{2}
\ldots \sum_{k=3,k\neq K}^{K}|u_{k}|^{2})\Big)$, whereas the first
term provides the condition $\nu^{2} |A_{1}|^{2}|\hat{\bf
h}_{1}^{H}{\bf h}_{1}|^{2} \geq 1/4$ that ensures the convexity of
$\tilde{J}_{CM}(.)$ in the noiseless case. Because $\bar{\bf
u}={\bf B}{\bf w}$ is a linear function of ${\bf w}$ then
$\tilde{J}_{CM}(\bar{\bf u})$ being a convex function of $\bar{\bf
u}$ implies that ${J}_{CM}({\bf w})=\tilde{J}_{CM}({\bf B}{\bf
w})$ is a convex function of ${\bf w}$. Since the extrema of the
cost function can be considered for small $\sigma^{2}$ a slight
perturbation of the noise-free case \cite{kwak}, the cost function
is also convex for small $\sigma^{2}$ when $\nu^{2}
|A_{1}|^{2}|\hat{\bf h}_{1}^{H}{\bf h}_{1}|^{2} \geq 1/4$. If we
assume ideal channel estimation ($|\hat{\bf h}_{1}^{H}{\bf
h}_{1}|=1$) and $\nu=1$, our result reduces to $|A_{1}|^{2}\geq
1/4$, which is the same found in \cite{linde}. For larger values
of $\sigma^{2}$, we remark that the term $\nu$ can be adjusted in
order to make the cost function $J_{CM}$ in (10) convex, as
pointed out in \cite{kwak}.

\section{Derivation of Normalized Step Size: CCM-SG case}

To derive a normalized step size for the algorithm in (24) and
(25) let us drop the time index (i) for simplicity and write the
constant modulus cost function $J_{CM}=(|{\bf w}_{k}^{H}{\bf
r}-{\bf f}_{k}^{H}\hat{\bf b}|^{2}-1)^{2}$ as a function of (24)
and (25):
\begin{equation}
J_{CM} = (|{\bf P}_{k}({\bf w}_{k} - \mu_{w}{\bf
r}e_{k}z_{k}^{*})^{H}{\bf r} - {\bf f}_{k}^{H}\hat{\bf b} -
\mu_{f}e_{k}^{*}z_{k}\hat{\bf b}^{H}\hat{\bf b} + $$ $$ ({\bf
C}_{k}({\bf C}_{k}^{H}{\bf C}_{k})^{-1}{\bf h}_{k})^{H} {\bf
r}|^{2} -1 )^{2}
\end{equation}
If we substitute ${\bf P}_{k} = {\bf I} -({\bf C}_{k}({\bf
C}_{k}^{H}{\bf C}_{k})^{-1}{\bf C}^{H}_{k}$ into the first term of
(44) and use ${\bf C}_{k}^{H}{\bf w}_{k}={\bf h}_{k}$ we can
simplify (44) and obtain:
\begin{equation}
J_{CM} = (|z_{k} -\mu_{w}e_{k}z_{k}{\bf r}^{H}{\bf P}_{k}{\bf r} -
\mu_{f}e_{k}z_{k} \hat{\bf b}^{H}\hat{\bf b}|^{2} -1 )^{2}
\end{equation}
Next, if we take the gradient of $J_{CM}$ with respect to
$\mu_{w}$ and equal it to zero, we have:
\begin{equation}
\nabla J_{\mu_{w}} = 2(|z_{k} -\mu_{w}e_{k}z_{k}{\bf r}^{H}{\bf
P}_{k}{\bf r} - \mu_{f}e_{k}z_{k} \hat{\bf b}^{H}\hat{\bf b}|^{2}
-1 ) \times $$ $$ \frac{d}{d \mu_{w}}|z_{k} -\mu_{w}e_{k}z_{k}{\bf
r}^{H}{\bf P}_{k}{\bf r} - \mu_{f}e_{k}z_{k} \hat{\bf
b}^{H}\hat{\bf b}|^{2}=0
\end{equation}
From the above expression it is clear that this minimization leads
to four possible solutions, namely:
\begin{equation}
\mu_{w}^{n.1} = \mu_{w}^{n.2} = \frac{1-\mu_{f}e_{k}\hat{\bf
b}^{H}\hat{\bf b} }{e_{k}{\bf r}^{H}{\bf P}_{k}{\bf r}}, ~
\mu_{w}^{n.3} = \frac{(|z_{k}|-1)-\mu_{f}|z_{k}|e_{k}\hat{\bf
b}^{H}\hat{\bf b} }{|z_{k}|e_{k}{\bf r}^{H}{\bf P}_{k}{\bf r}}, $$
$$ \mu_{w}^{n.4} = \frac{(|z_{k}|+1)-\mu_{f}|z_{k}|e_{k}\hat{\bf
b}^{H}\hat{\bf b} }{|z_{k}|e_{k}{\bf r}^{H}{\bf P}_{k}{\bf r}}
\end{equation}
By computing the second derivative of (44) one can verify that it
is always positive for the third and fourth solutions above,
indicating the minimum point. It should be remarked that the
solution for $\mu_{f}$ is analogous to $\mu_{w}$ and leads to the
same relations. Hence, we choose $\mu_{w} =
\frac{(|z_{k}|+1)-\mu_{f}|z_{k}|e_{k}\hat{\bf b}^{H}\hat{\bf b}
}{|z_{k}|e_{k}{\bf r}^{H}{\bf P}_{k}{\bf r}}$ and introduce again
the convergence factors $\mu_{0_{w}}$ and $\mu_{0_{f}}$ so that
the algorithms can operate with adequate step sizes that are
usually small to ensure good performance, and thus we have
$\mu_{w} = \mu_{0_{w}} \frac{(|z_{k}|+1)- \mu_{f}|z_{k}|e_{k}
\hat{\bf b}^{H}\hat{\bf b} }{|z_{k}|e_{k}{\bf r}^{H}{\bf
P}_{k}{\bf r}}$ and $\mu_{f} = \mu_{0_{f}} \frac{(|z_{k}|+1)-
\mu_{w}|z_{k}|e_{k}{\bf r}^{H}{\bf P}_{k}{\bf r} }{|z_{k}| e_{k}
\hat{\bf b}^{H}\hat{\bf b}}$.

\section{Derivation of Normalized Step Size: CMV-SG case}

To derive a normalized step size for the SG algorithm in (26) and
(27) let us let us drop again the time index (i) for simplicity
and write the minimum variance cost function $J=|{\bf
w}_{k}^{H}{\bf r}-{\bf f}_{k}^{H}\hat{\bf b}|^{2}$ as:
\begin{equation}
J_{MV} = |{\bf P}_{k}({\bf w}_{k} - \mu_{w}{\bf
r}x_{k}^{*})^{H}{\bf r} - {\bf f}_{k}^{H}\hat{\bf b} -
\mu_{f}x_{k}\hat{\bf b}^{H}\hat{\bf b} + $$ $$ + {\bf C}_{k}({\bf
C}_{k}^{H}{\bf C}_{k})^{-1}{\bf h}_{k})^{H} {\bf r}|^{2}
\end{equation}
If we take the gradient of $J_{MV}$ with respect to $\mu_{w}$ and
equal it to zero, we get:
\begin{equation}
\nabla J_{\mu_{w}} = 2|{\bf P}_{k}({\bf w}_{k} - \mu_{w}{\bf
r}x_{k}^{*})^{H}{\bf r} - {\bf f}_{k}^{H}\hat{\bf b} -
\mu_{f}x_{k}\hat{\bf b}^{H}\hat{\bf b} + $$ $$ + {\bf C}_{k}({\bf
C}_{k}^{H}{\bf C}_{k})^{-1}{\bf h}_{k})^{H} {\bf r}|~ \times
(-{\bf P}_{k}{\bf r}{x}_{k}^{*})^{H} {\bf r}=0
\end{equation}
If we substitute ${\bf P}_{k} = {\bf I} -({\bf C}_{k}({\bf
C}_{k}^{H}{\bf C}_{k})^{-1}{\bf C}^{H}$ into the first term of
(49) and use ${\bf C}_{k}{\bf w}_{k}={\bf h}_{k}$ we can eliminate
the third term of (49) and obtain the solution:
\begin{equation}
\mu_{w} = \frac{x_{k}(1-\mu_{f}\hat{\bf b}^{H}\hat{\bf
b})}{x_{k}({\bf r}^{H}{\bf P}_{k}{\bf r})} =
\frac{(1-\mu_{f}\hat{\bf b}^{H}\hat{\bf b})}{{\bf
r}^{H}\boldsymbol{\Pi}_{k}{\bf r}}
\end{equation}
Note that we introduce again a convergence factor $\mu_{0_{w}}$ so
that the algorithm can operate with adequate step sizes that are
usually small to ensure good performance, and thus we have
$\mu_{w} = \mu_{0_{w}}\frac{(1-\mu_{f}\hat{\bf b}^{H}\hat{\bf
b})}{{\bf r}^{H}{\bf P}_{k}{\bf r}}$. Next, we take the gradient
of $J_{MV}$ with respect to $\mu_{f}$ and equal it to zero:
\begin{equation}
\nabla J_{\mu_{f}} = 2|{\bf P}_{k}({\bf w}_{k} - \mu_{w}{\bf
r}x_{k}^{*})^{H}{\bf r} - {\bf f}_{k}^{H}\hat{\bf b} -
\mu_{f}x_{k}\hat{\bf b}^{H}\hat{\bf b} + $$ $$ +{\bf C}_{k}({\bf
C}_{k}^{H}{\bf C}_{k})^{-1}{\bf h}_{k})^{H} {\bf r}|~ \times
(-{x}_{k}~\hat{\bf b}^{H}\hat{\bf b})^{H} {\bf r}=0
\end{equation}
where it is noticed that the conditions are the same as for
$\mu_{w}$. Thus we proceed similarly to obtain the step size
$\mu_{f}$, which is given by $ \mu_{f} = \frac{(1-\mu_{w}{\bf
r}^{H}{\bf P}_{k}{\bf r})}{\hat{\bf b}^{H}\hat{\bf b}}$. Remark
again that we a convergence factor $\mu_{0_{f}}$ is applied so
that the algorithm can operate with adequate step sizes that are
usually small to ensure good performance, and thus we employ
$\mu_{f} = \mu_{0_{f}}\frac{(1-\mu_{w}{\bf r}^{H}{\bf P}_{k}{\bf
r})}{\hat{\bf b}^{H}\hat{\bf b}}$.

\section{On the Use of ${\bf R}_{k}$ for Channel Estimation}

Here, we discuss the suitability of the matrix ${\bf R}_{k}$, that
arises from the CCM method, for use in channel estimation. From
the analysis in Appendix I for the linear receiver, we have for an
ideal and asymptotic case that $u_{k} = (A_{1}^{*}{\bf
s}_{k}^{H}){\bf w}_{1} \approx 0$, for $k=2,\ldots,K$. Then, ${\bf
w}^{H}_{1}{\bf r} \approx A_{1}b_{1}{\bf w}^{H}_{1}{\bf s}_{1} +
{\bf w}^{H}_{1}{\bf n}$ and $|{\bf w}^{H}_{1}{\bf r}|^{2} \approx
A_{1}^{2}|{\bf w}_{1}^{H}{\bf s}_{1}|^{2} + A_{1}b_{1}({\bf
w}_{1}{\bf s}_{1}){\bf n}^{H}{\bf w}_{1} + A_{1}b_{1}^{*}({\bf
s}_{1}^{H}{\bf w}_{1}){\bf w}_{1}^{H}{\bf n} + {\bf w}^{H}_{1}{\bf
n}{\bf n}^{H}{\bf w}_{1}$. Therefore, we have for the desired user
(i.e. user $1$):
\begin{equation}
\begin{split}
{\bf R}_{1} & =E[|{\bf w}^{H}_{1}{\bf r}|^{2}{\bf r}{\bf r}^{H}]
\\ & \cong A_{1}^{2} |{\bf w}_{1}^{H}{\bf s}_{1}|^{2} {\bf R} + A_{1}
{\bf w}_{1}^{H}{\bf s}_{1}E[ b_{1}{\bf n}^{H}{\bf w}_{1}{\bf
r}{\bf r}^{H}] + \\ & ~ A_{1} {\bf s}_{1}^{H}{\bf w}_{1}E[
b_{1}^{*}{\bf w}^{H}_{1}{\bf n}{\bf r}{\bf r}^{H}] + E[{\bf
w}^{H}_{1}{\bf n}{\bf n}^{H}{\bf w}_{1} {\bf n}{\bf n}^{H}] +
\sigma^2 {\bf Q}{\bf w}_{1}^{H}{\bf w}_{1}
\\
& \cong A_{1}^{2} |{\bf w}_{1}^{H}{\bf s}_{1}|^{2} {\bf R} + A_{1}
{\bf w}_{1}^{H}{\bf s}_{1}E[ b_{1}{\bf n}^{H}{\bf w}_{1}{\bf
r}{\bf r}^{H}] + E[|{\bf w}^{H}_{1}{\bf n}|^{2} {\bf n}{\bf
n}^{H}] \\ & + A_{1} {\bf s}_{1}^{H}{\bf w}_{1}E[ b_{1}^{*}{\bf
w}^{H}_{1}{\bf n}{\bf r}{\bf r}^{H}]  + \sigma^2 ({\bf
R}-\sigma^{2}{\bf I}){\bf w}_{1}^{H}{\bf w}_{1} \\
& \cong (A_{1}^{2} |{\bf w}_{1}^{H}{\bf s}_{1}|^{2} + \sigma^{2})
{\bf R} + A_{1}^{2}\sigma^{2} ({\bf w}_{1}^{H}{\bf s}_{1})({\bf
w}_{1}{\bf s}_{1}^{H}) \\ & +  A_{1}^{2} \sigma^{2}({\bf
s}_{1}^{H}{\bf w}_{1})({\bf s}_{1}{\bf w}_{1}^{H}) + \sigma^{4}
\big[ {\rm diag}(|w_{1}|^{2},\ldots,|w_{N}|^{2}) \\ & + {\bf
w}_{1}{\bf w}_{1}^{H}\big] - \sigma^{4}{\bf
w}_{1}^{H}{\bf w}_{1}{\bf I} \\
& \cong A_{1}^{4} \Bigg[ \Bigg(\frac{|{\bf w}_{1}^{H}{\bf
s}_{1}|^{2}}{A_{1}^{2}} + \frac{\sigma^{2}}{A_{1}^{2}}\Bigg) {\bf
R} + \frac{\sigma^{2}}{A_{1}^{2}} \Big( ({\bf w}_{1}^{H}{\bf
s}_{1})({\bf w}_{1}{\bf s}_{1}^{H}) \\ & + \sigma^{2}({\bf
s}_{1}^{H}{\bf w}_{1})({\bf s}_{1}{\bf w}_{1}^{H}) \Big) +
\frac{\sigma^{4}}{A_{1}^{4}} \Big( \big[ {\rm
diag}(|w_{1}|^{2},\ldots,|w_{N}|^{2})  \\ & + {\bf w}_{1}{\bf
w}_{1}^{H}\big] - {\bf w}_{1}^{H}{\bf w}_{1}{\bf I} \Big) \Bigg]
\\ & \cong \alpha {\bf R} + \tilde{\bf N}
\end{split}
\end{equation}
where ${\bf R} = {\bf Q} +\sigma^2{\bf I}$, ${\bf Q} = E[{\bf
x}{\bf x}^{H}] = \sum_{k=1}^{K}|A_{k}|^{2}{\bf s}_{k}{\bf
s}_{k}^{H}$. From (52), it can be seen that ${\bf R}_{k}$ can be
approximated by ${\bf R}$ multiplied by a scalar factor $\alpha$
plus a noise-like term $\tilde{\bf N}$, that for sufficient
$E_{b}/N_{0}$ has an insignificant contribution. In addition, when
the symbol estimates $z_{k}={\bf w}_{k}^{H}{\bf r}$ are reliable,
that is the cost function in (10) is small ($J_{CM} <<1$), then
$|z_{k}|^{2}$ has small variations around unity for both linear
and DF detectors (note that $z_{k}={\bf w}_{k}^{H}{\bf r} - {\bf
f}_{k}^{H}\hat{\bf b}$ for the DF receivers), yielding the
approximation
\begin{equation}
E[|z_{k}|^{2}{\bf r}{\bf r}^{H}] = E[{\bf r}{\bf r}^{H}] +
E[(|z_{k}^{2} - 1){\bf r}{\bf r}^{H}] \cong E[{\bf r}{\bf
r}^{H}]={\bf R}
\end{equation}
Therefore, we conclude that the channel estimation can be
performed on ${\bf R}_{k}$ in lieu of ${\bf R}$, since the
properties of the matrix ${\bf R}$ studied in \cite{douko1,douko2}
hold for ${\bf R}_{k}$.
\end{appendix}

\end{document}